\documentclass[letter]{aa}  

\usepackage{graphicx}
\usepackage{txfonts}
\newcommand{\mycomment}[1]{}

\begin{document}

\title{The occurrence rate of galaxies with polar structures may be significantly underestimated}

\titlerunning{The occurrence rate of polar structures}

\author{A.V. Mosenkov\inst{1}, S.K.H. Bahr\inst{1}, V.P. Reshetnikov\inst{2,3,4},
           Z. Shakespear\inst{1}
          \and
          D.V. Smirnov\inst{2}
          }

\institute{Department of Physics and Astronomy, N283 ESC, Brigham Young University, Provo, UT 84602, USA
         \and
St.Petersburg State University, Universitetskii pr. 28, St.Petersburg, 198504 Russia
         \and
Central (Pulkovo) Astronomical Observatory, Russian Academy of Sciences, Pulkovskoye Chaussee 65/1, 196140 St. Petersburg, Russia
         \and
Special Astrophysical Observatory, Russian Academy of Sciences, Nizhnii Arkhyz, 369167 Russia
             }

\date{Received November, 2023; accepted ???, 2023}

\abstract{Polar-ring galaxies are photometrically and kinematically decoupled systems that are highly inclined to the major axis of the host. These galaxies have been explored since the 1970s, but the rarity of these systems has made such systematic studies difficult. However, over 250 
good candidates have been identified. In this work, we examine a sample of over 18,000 galaxies from the Sloan Digital Sky Survey (SDSS) Stripe\,82 for the presence of galaxies with polar structures. Using deep SDSS Stripe\,82, DESI Legacy Imaging Surveys, and Hyper Suprime-Cam Subaru Strategic Program, we selected 53 good candidate galaxies with photometrically decoupled polar rings, 9 galaxies with polar halos, 6 galaxies with polar bulges, and 34 possibly forming polar-ring galaxies, versus 13 polar-ring candidates previously selected in Stripe\,82. Our results suggest that the occurrence rate of galaxies with polar structures may be significantly underestimated, as revealed by the deep observations, and may amount to 1-3\% of non-dwarf galaxies.}

\keywords{galaxies: structure -- galaxies: elliptical and lenticular -- galaxies: formation -- galaxies: evolution}

\maketitle
\nolinenumbers

%
%________________________________________________________________

\section{Introduction} 
\label{sec:intro}

Polar-ring galaxies (PRGs), first identified in a series of studies by \citet{1978AJ.....83.1360S}, \citet{1978ApJ...226L.115B}, and \citet{1983AJ.....88..909S}, are a rare type of galaxy characterised by a ring of stars and gas orbiting (almost) perpendicular to the plane of the host galaxy \citep[see also][]{1990AJ....100.1489W,1997A&A...325..933R,2011MNRAS.418..244M}. However, a more general class of polar or tilted structures can be proposed, which includes polar or tilted discs \citep{1999ApJ...519L.127B,2020MNRAS.497.2039M}, polar bulges \citep{2012MNRAS.423L..79C,2015AstL...41..748R}, outer halos arranged orthogonally to the disc \citep{2016ApJ...823...19C,2020MNRAS.494.1751M,2021MNRAS.506.5030M}, and bright polar or highly tilted loops consisting of the material of disrupted satellites orbiting close to the host galaxy \citep{2010AJ....140..962M,2019A&A...632L..13M,2023A&A...671A.141M}. This broad class of galaxies with polar structures may serve as an important tool to probe the dark matter distribution in galaxies \citep{1987ApJ...314..439W,1994ApJ...436..629S,1996A&A...305..763C,2003ApJ...585..730I,2011MNRAS.418..244M,2012MNRAS.425.1967S,2014MNRAS.441.2650K,2015BaltA..24...76M} and the dynamics of galaxy accretion \citep{2008ApJ...689..678B} and mergers \citep{1998ApJ...499..635B,2003A&A...401..817B}. Currently, PRGs are thought to be formed as the result of i) tidal accretion of matter from a donor galaxy onto the host \citep[][]{schweizer1983colliding,1997A&A...325..933R} including disruption of a (gas-rich) satellite in a plane orthogonal to the plane of the host \citep{1991wdir.conf..112R,1992ApJ...389L..55K}, ii) major mergers of galaxies \citep[][]{bekki1997formation,bekki1998formation,2003A&A...401..817B}, and iii) cold accretion onto the host via cosmological filaments \citep{2006ApJ...636L..25M,2008ApJ...689..678B}. Modern cosmological hydrodynamical simulations show that any of these formation mechanisms may be responsible for the observed polar structures (\citealt{2019MNRAS.485..464L,2021MNRAS.504.5702W,2023arXiv231018597S}). On the other hand, deep images of PRGs often find debris from victim galaxies and arc-like structures, but also reveal signs of tidal accretion from a donor galaxy, implying that one or more formation mechanisms may contribute to the creation of PRGs \citep[][]{2011MNRAS.412..208F,2012MNRAS.422.2386F,mosenkov2022unveiling}.

Past catalogues of PRGs have been limited by their photometric depth\footnote{Here and below, the photometric depth of an image is defined as a $3\sigma$ of the background averaged over multiple randomly selected $10\times10$~arcsec$^{2}$ boxes in the $r$ band.}. 
For example, based on the preliminary morphological classification of almost 900,000 galaxies made by the Galaxy Zoo volunteers \citep{2011MNRAS.410..166L}, \citet{2011MNRAS.418..244M} examined over 40,000 SDSS galaxy images with an average photometric depth of 26.5 mag~arcsec$^{-2}$ and selected 275 PRG and PRG-related candidates for their Sloan Polar Ring Catalogue (SPRC).
Deep imaging has the potential to reveal previously unidentifiable, very faint details in and around galaxies \citep[see e.g.][and references therein]{abraham2014ultra,2015MNRAS.446..120D,2023A&A...671A.141M}. \citet{mosenkov2022unveiling} used Sloan Digital Sky Survey (SDSS, \citealt{2000AJ....120.1579Y}) Stripe\,82 deep imaging (with an average photometric depth of $28.89 \pm 0.25$ mag\,arcsec$^{-2}$) to study 13 PRGs listed in the SPRC and concluded that PRGs have a rich diversity of low surface brightness features. 

Unfortunately, the low average surface brightness of many of these structures, especially when viewed face-on, makes identification difficult and has prevented researchers from placing robust constraints on the prevalence of these structures. \citet{1990AJ....100.1489W} inferred that 0.5\% of nearby S0 galaxies should have identifiable polar rings, and that when corrected for various selection effects, this fraction may increase up to 5\% of all S0 galaxies. \citet{reshetnikov2011polar} estimated that 0.17\% of nearby galaxies with absolute magnitudes in the range of $M_\mathrm{B}=-17$ to $-22$~mag may have polar rings; however, if projection effects are taken into account, this fraction amounts to $\sim0.4$\%. Recently, \citet{2022MNRAS.516.3692S} investigated the luminosity functions of 103 `best' candidates for PRGs selected from \citet{1990AJ....100.1489W}, the SPRC, and \citet{2019MNRAS.483.1470R}. They concluded that only $\sim0.01$\% of nearby galaxies with absolute magnitudes in the range $M_\mathrm{r}=-17$ to -22~mag have polar rings, but their frequency of occurrence increases significantly with redshift up to $z\sim1$.

In this article, we aim to continue the study by \citet{mosenkov2022unveiling} and the examination of Stripe\,82 data to identify additional, previously unrecognised galaxies with polar rings. Surprisingly, we have found a total of 102 galaxies with polar rings, polar bulges and polar halos, many more than previously spotted using regular imaging, demonstrating that PRGs are more common than expected. The quantitative details of these candidates and the catalogue of galaxies with polar structures in Stripe\,82 will be provided in a further publication. However, in this letter, we report on a significant underestimate of the occurrence rate of PRGs in the local Universe as compared to the literature, which can be revealed through modern deep (down to 29--30~mag\,arcsec$^{-2}$) optical observations.

The remainder of this letter is organised as follows. In Section\,\ref{sec:data}, we describe the procedures used to obtain our sample, as well as the observations exploited.
%In Section\,\ref{sec:class}, we describe our classification scheme of galaxies with polar structures.
In Section\,\ref{sec:properties}, we describe some of the general properties of the candidates selected for PRGs. We draw our conclusions, with some implications of our study in Section\,\ref{sec:conclusions}. Throughout this article, we use the \citet{2020A&A...641A...6P} $\Lambda$CDM
cosmology.

\section{Data and sample selection}
\label{sec:data}
To select galaxies with polar structures from SDSS Stripe\,82, we mainly use the catalogue of 16,908 galaxies created by \citet{2019MNRAS.486..390B} based on the SDSS DR7. Furthermore, we retrieved an additional sample of galaxies within Stripe\,82 using the SDSS DR16 \citep{2020ApJS..249....3A} database not listed in the \citet{2019MNRAS.486..390B} catalogue and then the combined sample was filtered with the following conditions: redshift $z<0.3$ (to exclude too distant unresolved galaxies), Petrosian magnitude $m_\mathrm{r,Petro}\leq17.77$~mag (the faint limit of the SDSS Legacy spectroscopic sample), and Petrosian radius $R_\mathrm{r,Petro} \geq 7$~arcsec (to select  galaxies with a sufficient angular extent and examine their global morphology).
To limit the number of unusual outliers, we also removed a number of objects with colour index $g-r > 1$.
In total, our final sample of unique objects for revision in search of galaxies with polar structures is made up of 18,362 objects.

Each galaxy in our sample was visually inspected using co-added images from three different sources: the IAC Stripe\,82 Legacy Project \citep{2016MNRAS.456.1359F}, DESI Legacy Imaging Surveys DR9 (DESI hereafter, \citealt{2019AJ....157..168D}), and Hyper Suprime-Cam Subaru Strategic Program DR3 (HSC-SSP hereafter, \citealt{2022PASJ...74..247A}). This is done to ensure that the identified features in one survey are not simply image artefacts, but are also seen on images from other surveys.
The SDSS Stripe\,82 images retrieved for our sample have an average photometric depth of 28.5~mag\,arcsec$^{-2}$, whereas for DESI and HSC-SSP, we measured it to be 28.4~mag\,arcsec$^{-2}$ and 29.5~mag\,arcsec$^{-2}$, respectively. We note that only 76\% of the examined galaxies have images in HSC-SSP.

We sorted our sample of candidate polar structures into the following main categories: polar rings, polar halos, polar bulges, and forming polar rings. This classification was done in a purely qualitative manner, based on the characteristic morphological features for each of the following subsamples.

\textbf{Polar rings} exhibit a symmetric annular or elongated structure oriented at a right (high) angle relative to the major axis of the host.

\textbf{Polar bulges} represent prolate bulges, elongated perpendicular (or at some significant angle) to the host's disc plane. Such a tilt of the bulge is best identified in edge-on galaxies \citep{2015AstL...41..748R} because otherwise they can be confused with bars and other non-spherically symmetric central components.

\textbf{Polar halos} demonstrate a smooth, relatively uniform oval structure tilted at a high angle with respect to the major axis of the host galaxy. On closer inspection, these structures may, in fact, be a ring or consist of multiple loops and stellar streams embedded in a smooth envelope of stars, similar to what is observed on a deep image of the Sombrero galaxy \citep{2021MNRAS.506.5030M}.

\textbf{Forming polar structures}, primarily in the form of arcs, are highly tilted tidal structures observed around massive galaxies, often accompanied by tidally disrupted satellites. These structures represent distinctive trails of stars, gas, and dust brought into orbit around the host galaxy and may potentially produce a ring-like or halo-like structure in a highly tilted plane relative to the plane of the major galaxy \citep[see examples in][]{mosenkov2022unveiling}.

We note that the results of this classification are preliminary because they may be subject to our erroneous interpretation of the observed structures, so a quantitative analysis of the selected objects by means of host+ring decomposition (similar to that in \citealt{mosenkov2022unveiling}) will be performed in a subsequent work. In this paper, this classification is provided only for reference, and we do not use this separation into sub-classes in the following sections. 
Here, we focus only on the statistics of polar rings or related objects, whereas in future paper, we will present an atlas of all selected candidates for galaxies with polar structures and explore their morphological features in detail.

The comparison of imagery from different surveys was indeed helpful in detecting faint polar ring candidates in distant galaxies, especially using HSC-SSP data, which has a better average seeing of 0.7~arcsec in the $r$ band and $\sim1$~$r$-mag deeper photometry than in the other two surveys. In total, our sample comprises 102 galaxies with different kinds of polar or tilted structures, among which we selected 53 candidates for PRGs, 6 candidates for galaxies with polar bulges, 9 galaxies with polar halos, and 34 galaxies with possibly forming polar-ring structures. Of 102 candidates, only 8 were catalogued in \citet{2011MNRAS.418..244M}. To investigate how the better photometric depth of the images may have affected our classification, we examined the selected galaxies using `regular' SDSS images from DR16. Only 23 of the 102 galaxies demonstrate prominent polar structures on non-deep images. This suggests that deep imaging is indeed vital for revealing faint polar structures. 

\begin{figure*}
    \centering
    \includegraphics[width = .87\paperwidth]{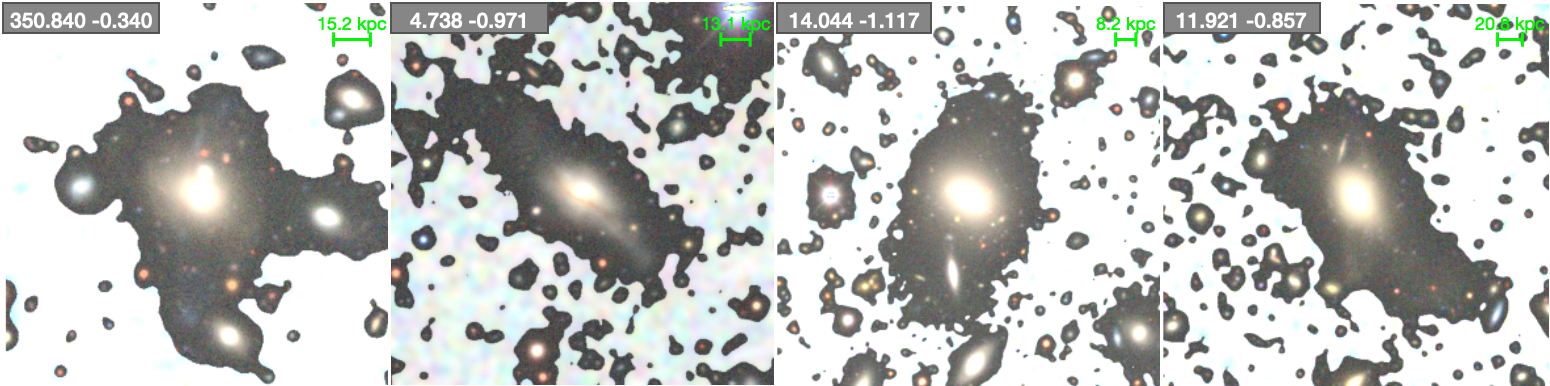}
    \caption{Examples of candidates for galaxies with polar structures from our sample as seen in the DESI Legacy from left to right: a polar-ring galaxy, a galaxy with a polar (tilted) bulge, a polar-halo galaxy, and a galaxy with a forming polar-ring structure. The coordinates for each galaxy are given in the upper left corner of each plot. The green bars depict a 10~arcsec scale. The darkish shadows show surface brightnesses brighter than 27~mag\,arcsec$^{-2}$ in the $r$ band.}
    \label{fig:examples}
\end{figure*}

In Fig.~\ref{fig:examples}, we present typical examples of candidates for galaxies with polar structures from the selected sample\footnote{Mosaic images for all candidates, along with their coordinates, are available in the supplementary material.}.

\section{Results and discussion}
\label{sec:properties}

\begin{figure*}
\centering
    \includegraphics[width = \linewidth]{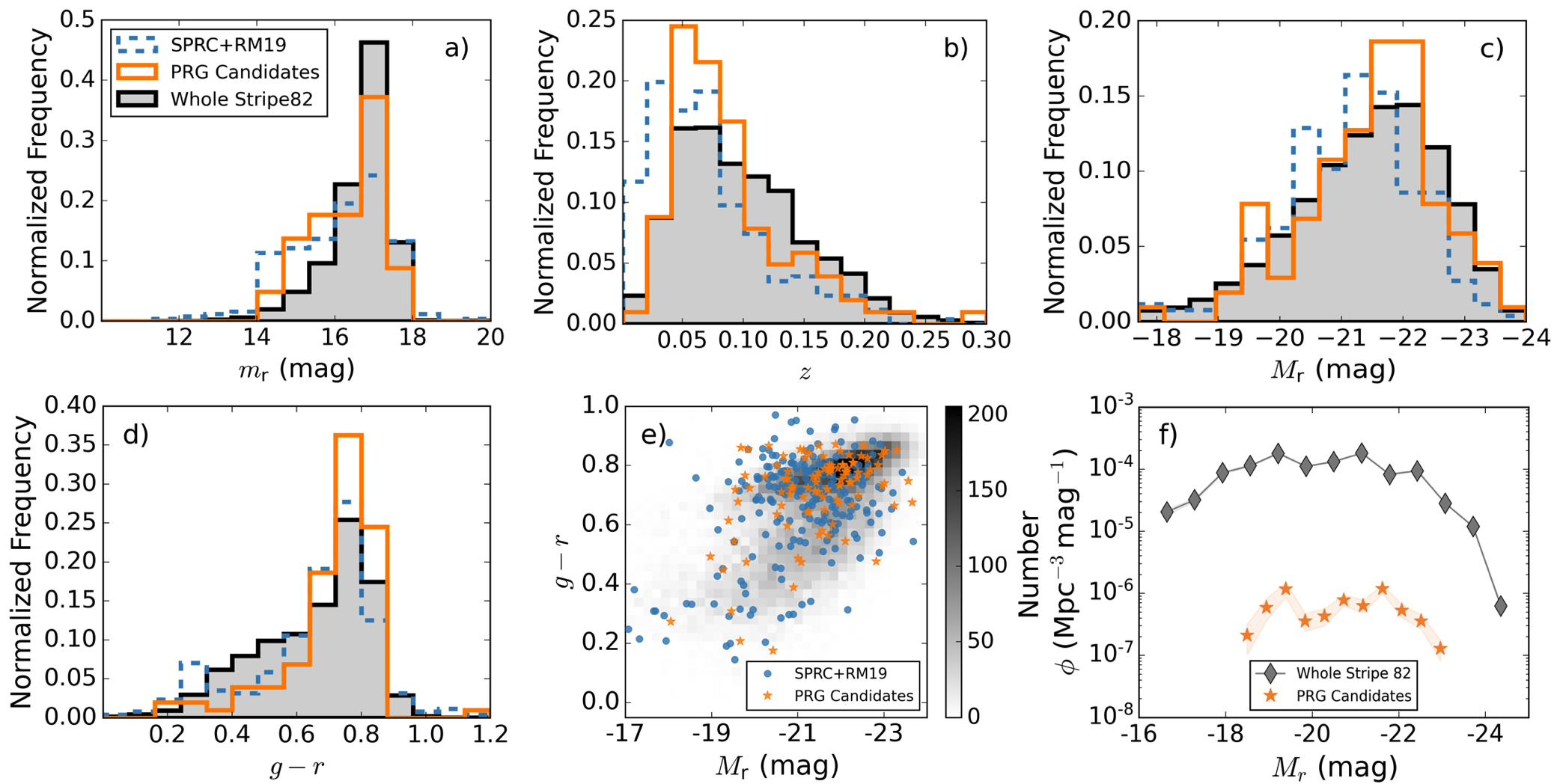}
    \caption{Normalized histograms comparing {\it a)} apparent magnitude in the $r$ band, {\it b)} redshift $z$, {\it c)} absolute magnitude in the $r$ band, and {\it d)} galaxy colour $g-r$. The black filled histograms represent the entire Stripe\,82 sample, while the SPRC+RM19 (see text)
    and our new candidates are shown as blue dashed and orange solid histograms, respectively. Plot {\it e)}: a colour-magnitude diagram (the underlying density plot depicts the distribution of the entire Stripe\,82 sample of the 18,362 galaxies). All magnitudes have been corrected for K-correction \citep{2010MNRAS.405.1409C,2012MNRAS.419.1727C} and Galactic extinction using the conversions from \citet{2011ApJ...737..103S} applied to $E(B-V)$ from \citet{1998ApJ...500..525S}. Plot {\it f)}: luminosity functions of our new PRG sample and the entire Stripe\,82 sample calculated with Cho\l{}oniewski's method \citep{1986MNRAS.223....1C}.}
    \label{fig:all_plots}
\end{figure*}

Combining polar rings, polar halos, polar bulges (which 
are morphologically similar to PRGs), and forming polar rings, we found 102 candidates for PRGs. We included polar halos in our sample because some of them may be polar rings, but because of poor resolution, we may observe them as smooth oval orthogonal structures instead. In Fig.~\ref{fig:all_plots}{\it a--d}, we compare the properties of these PRG candidates with the entire Stripe\,82 sample of galaxies, as well as with the sample of PRGs found in the SPRC \citep[][only `good' and `best' candidates were selected, comprising 185 galaxies in total]{2011MNRAS.418..244M} and 31 PRG candidates from \citet{2019MNRAS.483.1470R}. We combined them altogether as the SPRC+RM19 sample.

When comparing our new candidates with the entire Stripe\,82 sample of 18,362 galaxies, we notice that the new PRG candidates tend to have slightly brighter apparent magnitudes and lower redshifts, mirroring the typical selection bias. However, the redshift distribution for our selected candidates drops at $z<0.05$ compared to that for the SPRC+RM19 sample, which can be attributed to the different selection criteria used for creating these samples. The SPRC+RM19 sample undercounts the occurrence rate of PRGs at higher redshifts because it was based on shallower SDSS data. For the same reason, it also overcounts their fraction at lower redshift, where PRGs can be easily identified. We also note that the selected PRG candidates tend to have red colours $0.6\lesssim g-r \lesssim0.8$, which indicates that the vast majority of the new candidates are old, early-type galaxies, a common characteristic of PRGs \citep[see e.g.][]{1990AJ....100.1489W}. This is supported by the average S\'ersic index (retrieved from \citealt{2019MNRAS.486..390B}) of our candidates: $3.5^{+1.9}_{-1.3}$. 
In general, the properties of the PS candidates are similar to those of the combined SPRC+RM19 sample, with a slight shift toward visibly fainter objects (the advantage of using deep surveys is evidently important in recognising polar structures in fainter objects) at slightly higher redshifts but, at the same time, somewhat higher luminosities. 
The colour-magnitude diagram in Fig.~\ref{fig:all_plots}{\it e} illustrates that
the bulk of our newly found PRG candidates have very similar red colours
and are located along the red sequence, while some fall in the green valley and a few galaxies in the blue cloud. Our sample has a lower fraction of late-type galaxies as compared to the SPRC+RM19 sample, which can again be attributed to the selection effect; in particular: blue, less luminous polar rings are easier to identify around lower-redshift galaxies than at higher redshift and our sample has a lower fraction of low-redshift galaxies than the SPRC+RM19 sample. Characterising galaxies with polar structures and their incidence rate is crucial for modeling galaxies in general because they serve as good probes for assessing our ability to reproduce the observable Universe using modern cosmological hydrodynamical simulations.

The occurrence rate of polar-ring galaxies in Stripe 82 was estimated using the approach described in \citet{2022MNRAS.516.3692S}. Initially, the completeness of the PRG and the entire Stripe\,82 samples was assessed using the $\left<V/V_\mathrm{max}\right>$ method \citep{1973ApJ...186..433H}, resulting in correction factors of 2.32 and 1.41, respectively. Subsequently, luminosity functions for both samples, shown in Fig.~\ref{fig:all_plots}{\it f}, were calculated using Cho\l{}oniewski's method \citep{1986MNRAS.223....1C}. By integrating the luminosity function across the observed range of absolute magnitudes, spatial density estimates of 4.8$\times 10^{-6}$\,Mpc$^{-3}$ for PRGs and 4.2$\times 10^{-4}$\,Mpc$^{-3}$ for all Stripe\,82 galaxies were obtained. The ratio of these values yields a PRG occurrence rate of 1.1\%, which increases to approximately 3\% when the projection effect is taken into account.
This is significantly higher than the previous estimates, for example, 0.4\% from \citet{reshetnikov2011polar}. Before this study, only 13 PRG candidates had been known in Stripe\,82 \citep{2011MNRAS.418..244M,mosenkov2022unveiling}, but this survey revealed at least seven times as many possible PRGs. This suggests that more polar rings are expected to become visible at greater photometric depths. This conclusion is in line with the recent study by \citet{2023MNRAS.525.4663D} who revealed two orthogonal H{\sc i} components in NGC\,4632 and NGC\,6156 in the framework of the ambitious WALLABY project. At least for NGC\,4632, the gaseous ring also has a very faint stellar counterpart (see their Fig.~5). The authors estimate the incidence rate of such galaxies in the WALLABY survey to be about 1-3\%, which suggests that hundreds of new PRGs may be discovered in the near future. It is quite likely that similarly to NGC\,4632, many gaseous polar rings may have dim stellar counterparts, although purely gaseous PRGs, devoid of a stellar orthogonal constituent, have also been found in observations \citep{2006A&A...451...99B,2009ApJ...696L...6S}.
Moreover, in a pilot study using the IllustrisTNG50 simulation \citep{2018MNRAS.473.4077P}, \citet{2023arXiv231018597S} selected 6 PRGs out of 1600 sufficiently massive galaxies at $z=0.05$, resulting in a PRG fraction of 0.4\% (lower than what we report here). However, they also identified several candidates for PRGs with a gaseous polar ring only, without a prominent stellar counterpart. It is of great interest for a future study to compare the properties of the observed low surface brightness polar structures selected in this study with those from the TNG simulation and explore the dominant mechanisms for the formation of these unique objects.

\section{Conclusions}
\label{sec:conclusions}

In this letter, we present a review of a sample of over 18,000 galaxies within the SDSS Stripe\,82 sample to search for signs of polar-ring structures. The photometric depth of Stripe\,82, DESI, and HSC-SSP imaging allowed us to probe polar structures at a depth of $\sim29-30$~ mag\,arcsec$^{-2}$ in the $r$ band, which has never been achieved before at the survey level. This search has allowed us to identify 102 galaxies with possible polar structures (53 candidates to PRGs, 9 galaxies with polar halos, 6 galaxies with polar bulges, and 34 forming PRGs). We also found over 50 other galaxies with possible polar features, such as polar bulges and polar tidal structures in the form of stellar streams, which may become morphologically similar to PRG galaxies after the disruption of the dwarf galaxy (see also examples of such galaxies in the Stellar Stream Legacy Survey; e.g. Figs. 3 and 4 in \citealt{2023A&A...671A.141M}).

Our sample implies an occurrence rate of 1.1\% (up to 3\%, when the effect of inclination is taken into account) for galaxies with polar structures, higher than any previous estimate (\citealt{1990AJ....100.1489W}, \citealt{2011MNRAS.418..244M}). This suggests that polar structures are more common than previously expected, with the discrepancy stemming from the photometric depth of the observations. We stress, however, that these galaxies are only candidates for polar structures and are yet to be kinematically confirmed. In a future study, we will release a full catalogue of our polar structure candidates, alongside photometric decompositions of some candidate galaxies identified here. 
This will allow for a more in-depth study of this intriguing and still quite rare type of peculiar galaxies.

\begin{acknowledgements}
Funding for the Sloan Digital Sky Survey IV has been provided by the Alfred P. Sloan Foundation, the U.S. Department of Energy Office of Science, and the Participating Institutions. SDSS-IV acknowledges
support and resources from the Center for High-Performance Computing at the University of Utah. The SDSS website is www.sdss.org.

SDSS-IV is managed by the Astrophysical Research Consortium for the
Participating Institutions of the SDSS Collaboration including the
Brazilian Participation Group, the Carnegie Institution for Science,
Carnegie Mellon University, the Chilean Participation Group, the French Participation Group, Harvard-Smithsonian Center for Astrophysics,
Instituto de Astrof\'isica de Canarias, The Johns Hopkins University, Kavli Institute for the Physics and Mathematics of the Universe (IPMU) /
University of Tokyo, the Korean Participation Group, Lawrence Berkeley National Laboratory,
Leibniz Institut f\"ur Astrophysik Potsdam (AIP),
Max-Planck-Institut f\"ur Astronomie (MPIA Heidelberg),
Max-Planck-Institut f\"ur Astrophysik (MPA Garching),
Max-Planck-Institut f\"ur Extraterrestrische Physik (MPE),
National Astronomical Observatories of China, New Mexico State University,
New York University, University of Notre Dame,
Observat\'ario Nacional / MCTI, The Ohio State University,
Pennsylvania State University, Shanghai Astronomical Observatory,
United Kingdom Participation Group,
Universidad Nacional Aut\'onoma de M\'exico, University of Arizona,
University of Colorado Boulder, University of Oxford, University of Portsmouth,
University of Utah, University of Virginia, University of Washington, University of Wisconsin,
Vanderbilt University, and Yale University.

The Legacy Surveys consist of three individual and complementary projects: the Dark Energy Camera Legacy Survey (DECaLS; NOAO Proposal ID \# 2014B-0404; PIs: David Schlegel and Arjun Dey), the Beijing-Arizona Sky Survey (BASS; NOAO Proposal ID \# 2015A-0801; PIs: Zhou Xu and Xiaohui Fan), and the Mayall z-band Legacy Survey (MzLS; NOAO Proposal ID \# 2016A-0453; PI: Arjun Dey). DECaLS, BASS and MzLS together include data obtained, respectively, at the Blanco telescope, Cerro Tololo Inter-American Observatory, National Optical Astronomy Observatory (NOAO); the Bok telescope, Steward Observatory, University of Arizona; and the Mayall telescope, Kitt Peak National Observatory, NOAO. The Legacy Surveys project is honored to be permitted to conduct astronomical research on Iolkam Du'ag (Kitt Peak), a mountain with particular significance to the Tohono O'odham Nation.

NOAO is operated by the Association of Universities for Research in Astronomy (AURA) under a cooperative agreement with the National Science Foundation.

This project used data obtained with the Dark Energy Camera (DECam), which was constructed by the Dark Energy Survey (DES) collaboration. Funding for the DES Projects has been provided by the U.S. Department of Energy, the U.S. National Science Foundation, the Ministry of Science and Education of Spain, the Science and Technology Facilities Council of the United Kingdom, the Higher Education Funding Council for England, the National Center for Supercomputing Applications at the University of Illinois at Urbana-Champaign, the Kavli Institute of Cosmological Physics at the University of Chicago, Center for Cosmology and Astro-Particle Physics at the Ohio State University, the Mitchell Institute for Fundamental Physics and Astronomy at Texas A\&M University, Financiadora de Estudos e Projetos, Fundacao Carlos Chagas Filho de Amparo, Financiadora de Estudos e Projetos, Fundacao Carlos Chagas Filho de Amparo a Pesquisa do Estado do Rio de Janeiro, Conselho Nacional de Desenvolvimento Cientifico e Tecnologico and the Ministerio da Ciencia, Tecnologia e Inovacao, the Deutsche Forschungsgemeinschaft and the Collaborating Institutions in the Dark Energy Survey. The Collaborating Institutions are Argonne National Laboratory, the University of California at Santa Cruz, the University of Cambridge, Centro de Investigaciones Energeticas, Medioambientales y Tecnologicas-Madrid, the University of Chicago, University College London, the DES-Brazil Consortium, the University of Edinburgh, the Eidgenossische Technische Hochschule (ETH) Zurich, Fermi National Accelerator Laboratory, the University of Illinois at Urbana-Champaign, the Institut de Ciencies de l'Espai (IEEC/CSIC), the Institut de Fisica d'Altes Energies, Lawrence Berkeley National Laboratory, the Ludwig-Maximilians Universitat Munchen and the associated Excellence Cluster Universe, the University of Michigan, the National Optical Astronomy Observatory, the University of Nottingham, the Ohio State University, the University of Pennsylvania, the University of Portsmouth, SLAC National Accelerator Laboratory, Stanford University, the University of Sussex, and Texas A\&M University.

BASS is a key project of the Telescope Access Program (TAP), which has been funded by the National Astronomical Observatories of China, the Chinese Academy of Sciences (the Strategic Priority Research Program "The Emergence of Cosmological Structures" Grant \# XDB09000000), and the Special Fund for Astronomy from the Ministry of Finance. The BASS is also supported by the External Cooperation Program of Chinese Academy of Sciences (Grant \# 114A11KYSB20160057), and Chinese National Natural Science Foundation (Grant \# 11433005).

The Legacy Survey team makes use of data products from the Near-Earth Object Wide-field Infrared Survey Explorer (NEOWISE), which is a project of the Jet Propulsion Laboratory/California Institute of Technology. NEOWISE is funded by the National Aeronautics and Space Administration.

The Legacy Surveys imaging of the DESI footprint is supported by the Director, Office of Science, Office of High Energy Physics of the U.S. Department of Energy under Contract No. DE-AC02-05CH1123, by the National Energy Research Scientific Computing Center, a DOE Office of Science User Facility under the same contract; and by the U.S. National Science Foundation, Division of Astronomical Sciences under Contract No. AST-0950945 to NOAO.

The Hyper Suprime-Cam (HSC) collaboration includes the astronomical communities of Japan and Taiwan, and Princeton University. The HSC instrumentation and software were developed by the National Astronomical Observatory of Japan (NAOJ), the Kavli Institute for the Physics and Mathematics of the Universe (Kavli IPMU), the University of Tokyo, the High Energy Accelerator Research Organization (KEK), the Academia Sinica Institute for Astronomy and Astrophysics in Taiwan (ASIAA), and Princeton University. Funding was contributed by the FIRST program from the Japanese Cabinet Office, the Ministry of Education, Culture, Sports, Science and Technology (MEXT), the Japan Society for the Promotion of Science (JSPS), Japan Science and Technology Agency (JST), the Toray Science Foundation, NAOJ, Kavli IPMU, KEK, ASIAA, and Princeton University.

This paper makes use of software developed for Vera C. Rubin Observatory. We thank the Rubin Observatory for making their code available as free software at http://pipelines.lsst.io/.

This paper is based on data collected at the Subaru Telescope and retrieved from the HSC data archive system, which is operated by the Subaru Telescope and Astronomy Data Center (ADC) at NAOJ. Data analysis was in part carried out with the cooperation of Center for Computational Astrophysics (CfCA), NAOJ. We are honored and grateful for the opportunity of observing the Universe from Maunakea, which has the cultural, historical, and natural significance in Hawaii.

\end{acknowledgements}

\bibliographystyle{aa} % style aa.bst
\bibliography{art}

%%%%%%%%%%%%%%%%%%%%%%%%%%%%%%%%%%%%%%%%%%%%%%%%%%

%%%%%%%%%%%%%%%%%%%%%%%%%%%%%%%%%%%%%%%%%%%%%%%%%%

%%%%%%%%%%%%%%% APPENDICES %%%%%%%%%%%%%%%%%%%%%

\end{document}